\documentclass[a4paper,fleqn]{cas-sc}
\usepackage{soul}
\usepackage[sort&compress,numbers]{natbib}
\usepackage{lineno}
\usepackage{setspace}
\usepackage{verbatim}
\usepackage{graphics}
\usepackage{graphicx}
\usepackage{cuted}
\usepackage{lipsum}
\usepackage{multirow}

\def\tsc#1{\csdef{#1}{\textsc{\lowercase{#1}}\xspace}}
\tsc{WGM}
\tsc{QE}
\tsc{EP}
\tsc{PMS}
\tsc{BEC}
\tsc{DE}

\begin{document}

\let\WriteBookmarks\relax
\def\floatpagepagefraction{1}
\def\textpagefraction{.001}
\shorttitle{Thermomechanical Stability of Porous Graphene}
\shortauthors{Pereira J\'unior \textit{et~al}.}

\title [mode = title]{Thermomechanical Insight into the Stability of Nanoporous Graphene Membranes}

\author[1]{Marcelo Lopes Pereira J\'unior}
\author[1,2]{Luiz Ant\^onio Ribeiro J\'unior}
\cormark[1]
\ead{ribeirojr@unb.br}

\address[1]{Institute of Physics, University of Bras\'ilia, 70910-900, Bras\'ilia, Brazil}
\address[2]{PPGCIMA, Campus Planaltina, University of Bras\'{i}lia, 73345-010, Bras\'{i}lia, Brazil}
\cortext[cor1]{Corresponding author}

\begin{abstract}
Porous graphene (PG) is a graphene derivative endowed of nanoporous architectures. This material possesses a particular structure with interconnected networks of high pore volume, producing membranes with a large surface area. Experiments revealed that PG combines remarkable properties such as high mechanical strength and good thermal stability. In this work, we have carried out fully-atomistic reactive (ReaxFF) molecular dynamics simulations to perform a comprehensive study on the elastic properties, fracture mechanism, and thermal stability of 2D porous $n$-Benzo-CMPs (CMP and $n$ refer, respectively, to $\pi$-conjugated microporous polymers and the pore diameter) membranes with distinct nanoporous architectures. For comparison purposes, the results were also contrasted with the ones for graphene sheets of similar dimensions. We adopted three different nanoporous diameters: small (3.45 \AA), medium (8.07 \AA), and large (11.93 \AA). Results showed that PG is thermally stable up to 4660K, about 1000K smaller than the graphene melting point (5643K). During the PG heating, linear atomic chains are formed combining carbon and hydrogen atoms. The fracture strains range between 15\%--34\% by applying a uniaxial loading in both plane directions for temperatures up to 1200K. The fracture strain increases proportionally with the nanoporous size. Remarkably, the critical tensile strength for the PG complete fracture is temperature independent. Instead, it depends only on the nanoporous diameter. All the PG membranes go abruptly from elastic to completely fractured regimes after a critical strain threshold.
\end{abstract}



\begin{keywords}
Reactive Molecular Dynamics \sep Porous Graphene Membranes \sep Elastic Properties \sep Fracture Pattern \sep Melting Point 
\end{keywords}

\maketitle
\doublespacing

\section{Introduction}

Porous Graphene (PG) membranes are graphene-related materials composed of a large surface area containing nanopores with different shapes and sizes \cite{zhang_SMALL,hooch_AMI}. PG membranes are interesting materials due to their low density and high strength, being capable of bonding with other atoms through their sp, sp$^{2}$, and sp$^{3}$ hybrid orbitals \cite{guirguis_CARBON,guirguis_MH}. They have emerged to overcome the problem of the severely reduced surface area presented by graphene layers when in contact, which heavily weakens several of their outstanding properties \cite{lokhande_JMCA}. In this sense, the surface area of graphene --- or assembly of graphene layers --- is enlarged by tailoring the lattice morphology via pore production \cite{jiang_NL}. PG membranes have been employed to develop several applications, ranging from biomedical \cite{olszowska_SM,jakus_ACSNANO} to energy storage and conversion devices \cite{sun_JMCA,zhang_JMR,ferrer_NANOCONV,yan_AFM,xiao_NL}. To further explore the functionalities of PG, many studies have been devoted to designing novel advanced PG-based materials with both 2D and 3D architectures and endowed of more accessible electroactive areas \cite{jiang_NS,meng_JMCA,du_JPCC}.         

PG membranes present a random or high regularity distribution of nanoporous \cite{moreno_SCIENCE}. These pores are classified according their diameters (D) as: micropores (D $<$ 2 nm), mesopores (2 nm $<$ D $<$ 50 nm), and macropores (D $>$ 50 nm) \cite{russo_NML}. To produce them, some sp$^2$ carbon atoms are removed from the pristine graphene lattice. Particularly, PG layers with micropores are employed in developing new technologies for water and air purification, due to its spongy structure \cite{koenig_NATNANOTECH}. The ones displaying mesopores and macropores, in turn, have been used as active layers in electrochemical capacitors \cite{zhou_JMCA}. Among all the species of 2D porous graphitic materials, $n$-Benzo-CMPs (where CMP and $n$ refer, respectively, to $\pi$-conjugated microporous polymers and the pore size), stand out due to their direct semiconducting bandgaps (0.6--1.75 eV), making them interesting for a wide range of optoelectronic applications \cite{li_JMCC}. Importantly, their synthesis has revealed novel functions for potential applications as super-capacitor in the energy storage or electric power supply \cite{xu_CSR}.       

Recently, several theoretical studies were performed inspired by the successful synthesis of 2D porous CMPs \cite{li_JMCC,esfandiarpoor_SCIREP,liu_SSC,wang_PCCP,tao_ACSAMI,hankel_JPCC,brunetto_JPCC,debing_JPCC}. The electronic and transport properties of $n$-Benzo-CMPs and BN codoped derivatives were investigated using density-functional theory (DFT) and the non-equilibrium Green's function methods \cite{li_JMCC}. The results showed that these materials possess interesting structural, electronic, and transport properties that are notably different from those of the graphene. Their direct semiconducting bandgaps ranged from 0.19 to 2.0 eV depending on the pore size diameter, which has varied within the interval 3.68--12.93 \AA, and the doping degree, in the sense that BN codoping tends to reduce the bandgaps \cite{li_JMCC}. Molecular dynamics (MD) simulations were performed to study the gas separation performance of PG membranes with H-passivated nanopores \cite{esfandiarpoor_SCIREP,liu_SSC,wang_PCCP,debing_JPCC,hankel_JPCC}. Some models of PG systems with different sizes and shape were designed to obtain an efficient CO$_{2}$/H$_{2}$ \cite{esfandiarpoor_SCIREP} and CO$_{2}$/N$_{2}$ \cite{wang_PCCP} separation. MD simulation results showed that H-passivated membranes with a pore size of 3.75~\AA perform high selectivity and desirable permeability for CO2/H2 separation while smaller and larger pores demonstrated less permeability and selectivity, respectively \cite{esfandiarpoor_SCIREP}. Moreover, it was reported in reference \cite{wang_PCCP} that H-passivated membranes with a pore size of 4.06~\AA can efficiently separate N$_2$ from CO$_2$. This particular PG membrane exhibited high N$_2$ selectivity over CO$_2$, with an N$_2$ permeance of 105 GPU (gas permeation unit) \cite{wang_PCCP}. No CO$_2$ was found to pass through the pore. The main conclusion in both studies is that the electrostatic sieving plays a fundamental role in hindering the passage of CO$_2$ molecules through the pore \cite{wang_PCCP}. As mentioned above, PG membranes have been successively employed to develop a wide range of applications. Nevertheless, some intrinsic thermomechanical properties of these materials, that may impact their performance in these applications, remain not described and further investigations are needed to fill this gap.  

In the present work, the mechanical and thermal stability of microporous 2D $n$-Benzo-CMPs membranes were systematically investigated in the scope of fully-atomistic reactive (ReaxFF) molecular dynamics simulations. We have considered three different H-passivated nanoporous diameters: small (3.45 \AA), medium (8.07 \AA), and large (11.93 \AA). A heating ramp protocol was applied to obtain the melting point for the PG membranes. Their mechanical behavior was studied using the stress-strain relationship and fracture toughness. 

\section{Methodology}

\begin{figure*}[pos=t]
	\centering
	\includegraphics[width=0.9\linewidth]{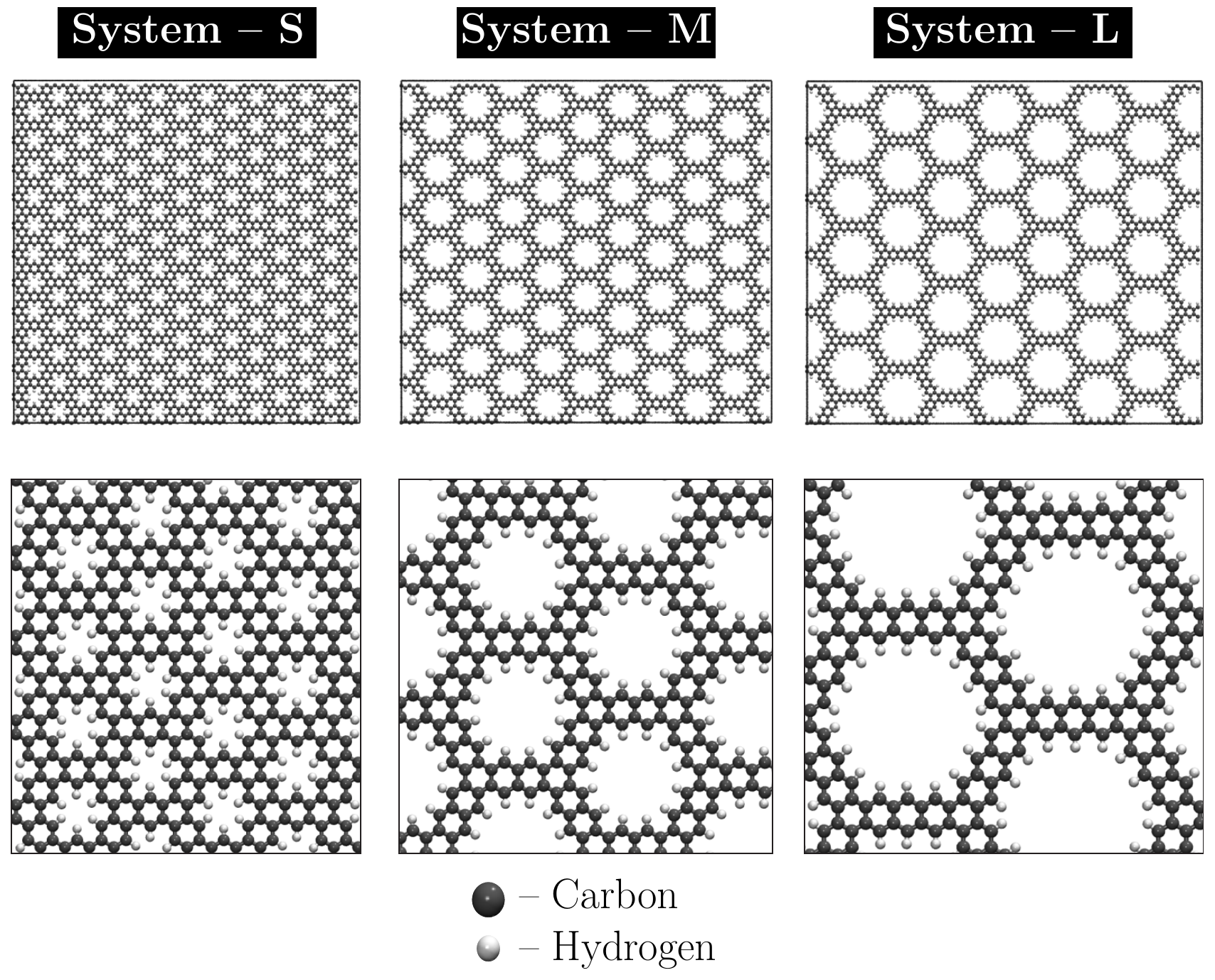} 
	\caption{Schematic representation of the microporous 2D $n$-Benzo-CMPs systems (porous graphene membranes) studied here. (a) $3.45$-Benzo-CMP (System -- S, 06 hydrogens in the pore), (b) $8.07$-Benzo-CMP (System -- M, 12 hydrogens in the pore), and $11.93$-Benzo-CMP (System -- L, 18 hydrogens in the pore). The carbon and hydrogen atoms are represented by the gray and white spheres, respectively. The system dimensions are: $102.13\times 100.81$ \AA${^2}$ (System -- S, 3024 carbons and 1008 hydrogens),  $109.5\times 100.9$ \AA${^2}$ (System -- M, 2400 carbons and 960 hydrogens), and  $116.87\times 100.79$ \AA${^2}$ (System -- L, 2016 carbons and 864 hydrogens).}
	\label{fig:membranes}
\end{figure*}

The mechanical and thermal properties of microporous 2D $n$-Benzo-CMPs membranes were studied through MD simulations using the reactive force field ReaxFF potential \cite{vanduin_JPCA,mueller_JPCC}, as implemented in the large-scale atomic/molecular massively parallel simulator (LAMMPS) code \cite{plimpton_JCP}. As expected in reactive potentials, the ReaxFF force field allows the formation and breaking of chemical bonds during the dynamics, which is useful to investigate the PG fracture mechanism. The studied H-passivated PG membranes are illustrated in Figure \ref{fig:membranes}. In this figure, it is depicted in the left-most panels the $3.45$-Benzo-CMP (System -- S, with 06 hydrogens in the pore), in the middle panels the $8.07$-Benzo-CMP (System -- M, with 12 hydrogens in the pore), and in the right-most panels the $11.93$-Benzo-CMP (System -- L, with 18 hydrogens in the pore). In Figure \ref{fig:membranes}, the carbon and hydrogen atoms are represented by the gray and white spheres, respectively. The system dimensions are: $102.13\times 100.81$ \AA${^2}$ (System -- S, 3024 carbons and 1008 hydrogens), $109.5\times 100.9$ \AA${^2}$ (System -- M, 2400 carbons and 960 hydrogens), and  $116.87\times 100.79$ \AA${^2}$ (System -- L, 2016 carbons and 864 hydrogens). It is worthwhile to stress that the systems were built in order to obtain 2D membranes with dimensions of about $100\times 100$ \AA${^2}$, preserving the periodic boundary conditions. 

The equations of motion were numerically integrated using the velocity-Verlet integrator with a time-step of $0.05$ fs. We increased the tensile stress in the system by applying a uniaxial strain along the periodic $x$ and $y$ directions, for an engineering strain rate of $10^{-6}$ fs$^{-1}$. The PG membranes were stressed up to their complete rupture, which is identified by the fractured and irreversible patterns presented by the systems when separated into two moieties. To eliminate any initial stress within the membranes, before the stretching procedure they were equilibrated/thermalized within an NPT/NVT ensembles at constant temperatures (300K, 600K, 900K, and 1200K) and null pressures using the Nos\'e-Hoover thermostat \cite{andersen_JCP,hoover_PRA} during 100 ps. Within this MD simulation protocol, Young’s modulus ($Y$), Fracture Strain ($FS$), and Ultimate Strength ($US$) are the elastic properties derived from the stress-strain curves obtained for each temperature considered here. 

To better analyze the outcomes from the stretching dynamics, we calculated the von Mises stress (VM) per-atom values \cite{mises_1913}. The VM values provide useful local structural information on the fracture mechanism, once they can determine the region from which the structure has started to yield the fractured lattice. In this way, the VM equation can be written as
\begin{equation}
\centering
\sigma^{k}_{v} = \sqrt{\frac{(\sigma^{k}_{xx} - \sigma^{k}_{yy})^2 + (\sigma^{k}_{yy} - \sigma^{k}_{zz})^2 + (\sigma^{k}_{xx} - \sigma^{k}_{zz})^2 + 6((\sigma^k_{xy})^2+(\sigma^k_{yz})^2+(\sigma^k_{zx})^2)}{2}},
\label{vm_equation}
\end{equation}
\noindent where $\sigma^k_{xx}$, $\sigma^k_{yy}$, and $\sigma^k_{zz}$ are the components of the normal stress and $\sigma^k_{xy}$, $\sigma^k_{yz}$, and $\sigma^k_{zx}$ are the components of the shear stress. The MD snapshots and trajectories were obtained by using free visualization and analysis software VMD \cite{HUMPHREY199633}.

\section{Results}

\begin{figure*}[pos=t]
	\centering
	\includegraphics[width=0.9\linewidth]{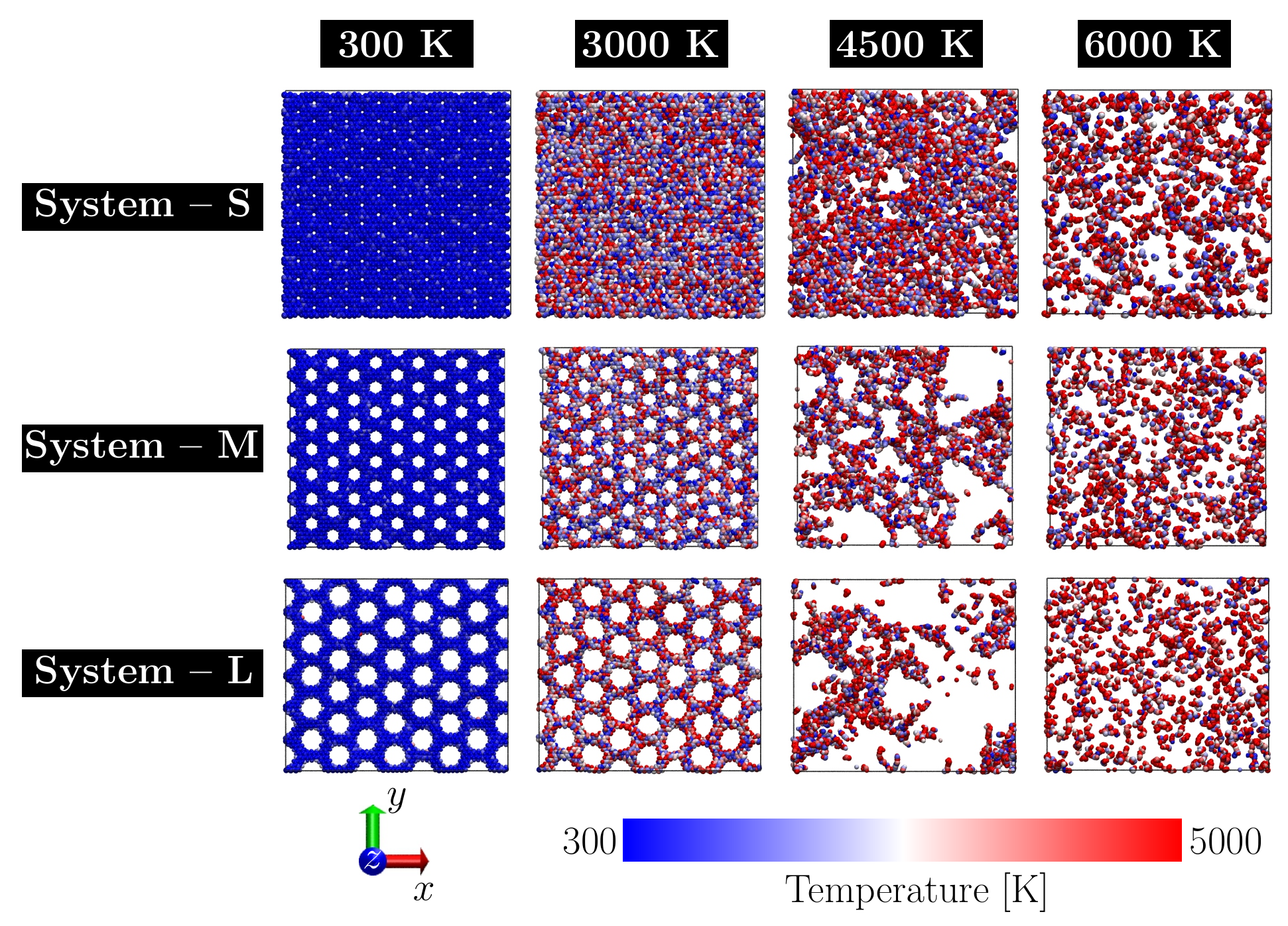} 
	\caption{Representative MD snapshots for the heating ramp simulations (melting process), with temperature varying from 300K up to 6000K for all the microporous 2D $n$-Benzo-CMPs systems studied here. The color scheme denotes the temperature per atom ranging from 300K to 5000K, which is presented by the blue and red colors, respectively.}
	\label{fig:mdmelting}
\end{figure*}

We begin our discussions by analyzing the thermal stability of the microporous 2D $n$-Benzo-CMPs systems studied here. Figure \ref{fig:mdmelting} shows representative MD snapshots for the heating ramp simulations (melting process), with temperature varying from 300K up to 6000K. In this figure, the columns present the lattice configurations for each kind of PG membrane at different temperatures, which increase from left to right. The color scheme denotes the temperature per-atom ranging from 300K to 5000K, which are represented by the blue and red colors, respectively. At low temperatures (300K), the PG membranes present lattice configurations very close to their ground state arrangement. For temperatures about 3000K, the thermal fluctuations lead the membranes to wrinkle due to large displacements presented by the atoms. Even so, the PG membranes keep their integrity. Above 4500K, the 2D $\pi$-conjugated polymer arrangement showed by these PG membranes favors the formation of very short and linear atomic chains (LACs), which randomly combine carbon and hydrogen atoms during the melting process. Importantly, the wrinkling effect also favors LACs formation. For temperatures about 6000K, only isolated atoms and small chain fragments that connect 2-10 atoms are observed. In this sense, the final morphologies obtained at the end of the melting process resemble a gas-like phase of carbon atoms.

\begin{figure*}[pos=t]
\centering
\includegraphics[width=0.9\linewidth]{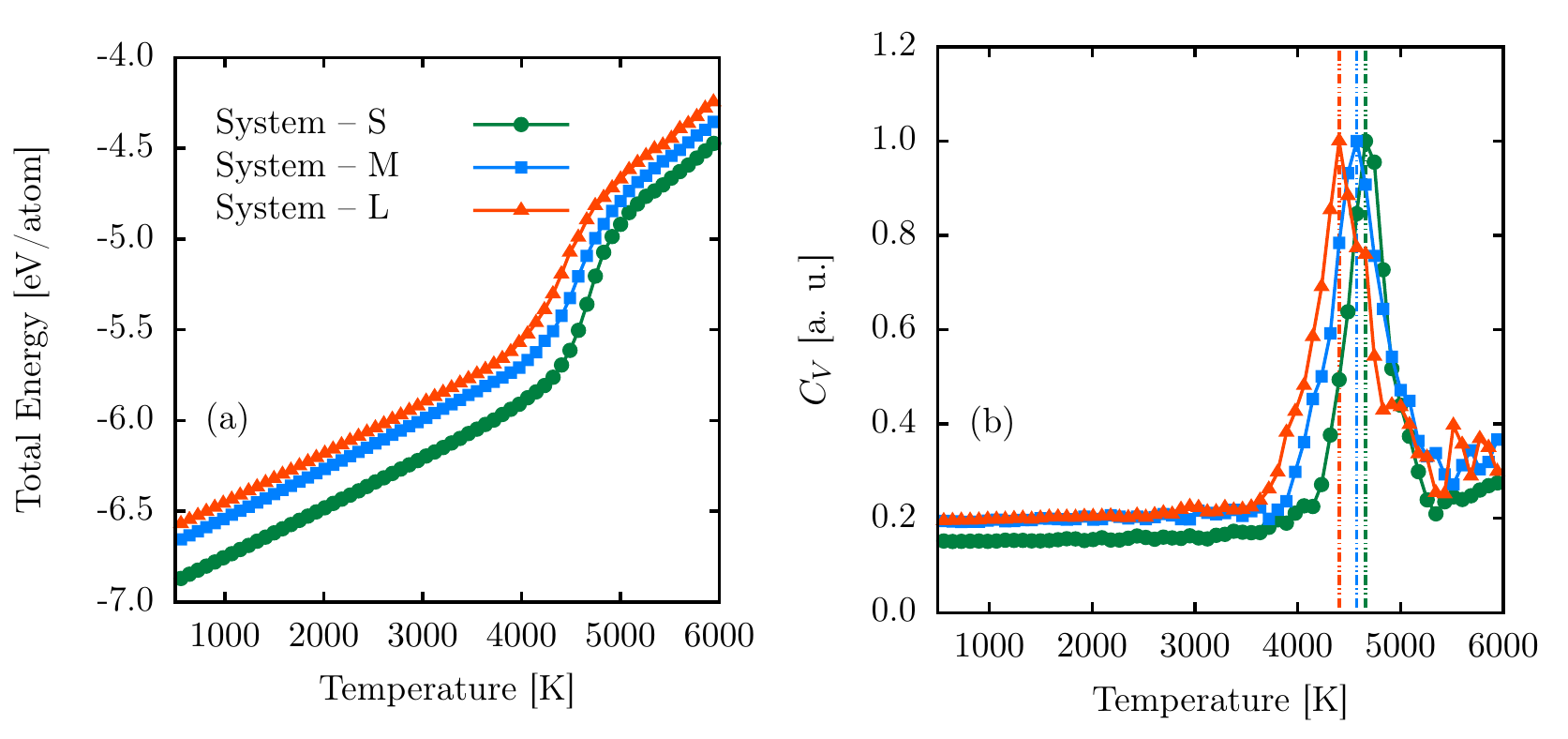} 
\caption{(a) Total energy and (b) heat capacity ($C_V$) as a function of temperature for the MD simulations presented in Figure \ref{fig:mdmelting}. In panel (b), the $C_V$ peaks (dashed lines) correspond precisely to the melting point of the PG membranes: 4660.07K, 4574.36K, and 4402.91 K, for System--S, System--M, and System--L, respectively.}
\label{fig:ramp}
\end{figure*}

Figure \ref{fig:ramp} depicts the total energy (\ref{fig:ramp}(a)) and the heat capacity ($C_V$, \ref{fig:ramp}(b)) as a function of temperature for the MD simulations presented in Figure \ref{fig:mdmelting}. In Figure \ref{fig:ramp}(a), one can note that the total energy increases linearly with temperature by showing two well-defined regimes with different slopes for all the systems. The curves are almost parallel and the total energy slightly increases by increasing the nanopore diameter. Moreover, we can see a clear discontinuity in these curves that denotes a phase transition from a solid to a gas-like phase. This abrupt change in the slope of the curves is related to a gain in kinetic energy due to the higher atom velocities in the gas-like phase, which increases the total energy. A considerable part of the harmonic and torsional energies present in the solid phase is converted into kinetic energy, increasing the amount of such a contribution for total energy. For the microporous 2D $n$-Benzo-CMPs systems considered here, the temperatures for the phase transitions occur at 4660.07K, 4574.36K, and 4402.91 K, for the systems S, M, and L, respectively. Importantly, a previous atomistic MD simulation has predicted a melting point for graphene about 5643K \cite{felix_JPCC}. As expected, when it comes to thermal stability, our results suggest that PG membranes with small nanopore diameters resemble more a graphene sheet than the ones with large nanopore diameters. In Figure \ref{fig:ramp}(b), the $C_V$ peaks (dashed lines) correspond precisely to the melting point of the PG membranes (mentioned above). This figure helps to realize that the critical temperature for a particular PG membrane (System--L) is about 1000K smaller than the graphene melting temperature \cite{felix_JPCC}, also pointing to the good thermal stability of these PG membranes species.  

\begin{figure*}[pos=t]
\centering
\includegraphics[width=0.9\linewidth]{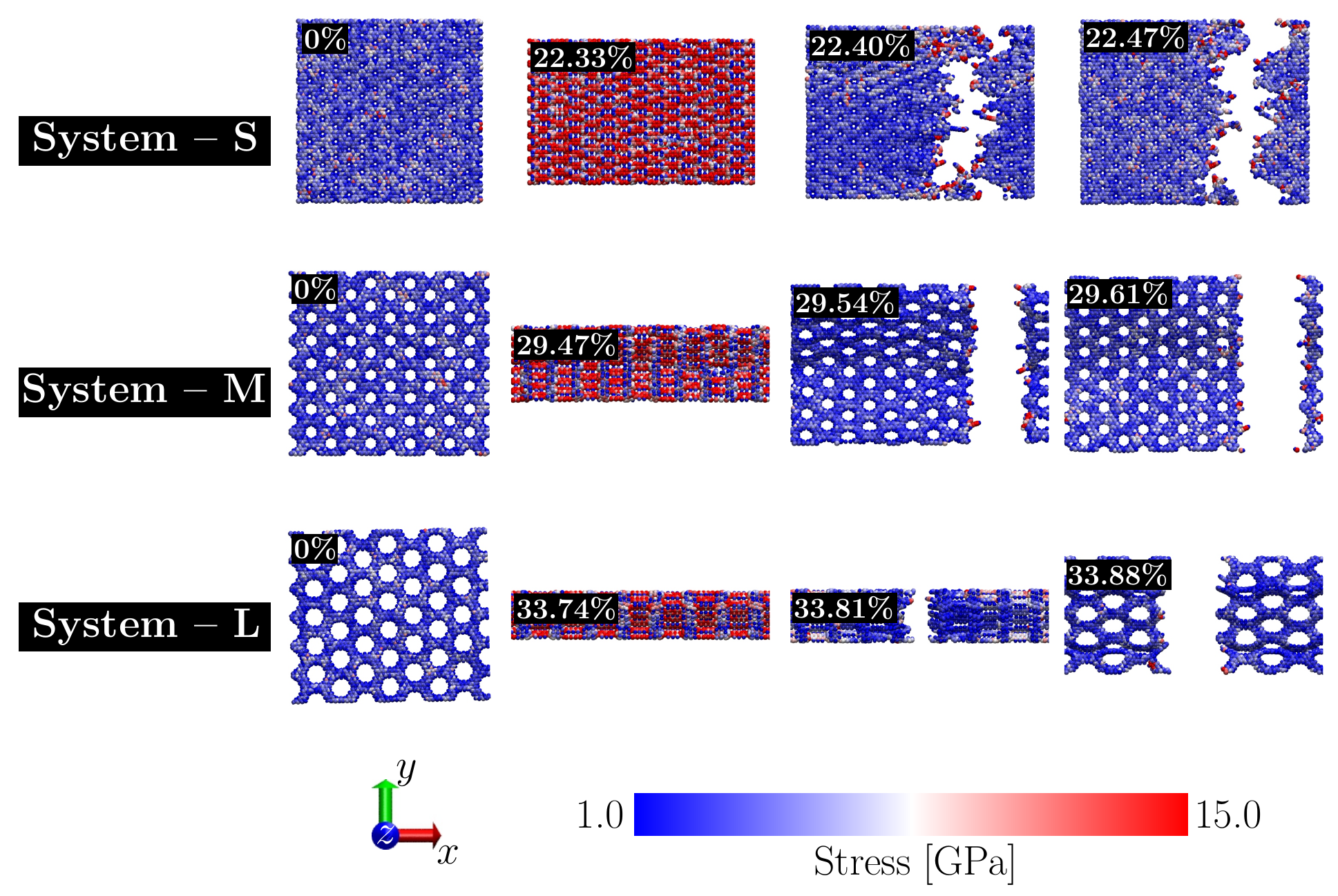} 
\caption{Representative MD snapshots of the strain simulations at 300K for all the microporous 2D $n$-Benzo-CMPs systems studied here. The color scheme refers to the minimum (blue) and maximum (red) VM stress values according to Equation \ref{vm_equation}.}
\label{fig:mdstress}
\end{figure*}

Now, we discuss the mechanical behavior of PG membranes under uniaxial ($x$-direction) tensile loading. Figure \ref{fig:mdstress} shows representative MD snapshots for strain simulations at 300K. In this figure, four different strain stages are illustrated for each system considered here. The color scheme refers to the minimum (blue) and maximum (red) VM stress values according to Equation \ref{vm_equation}. For the sake of convenience, here we not present the MD snapshots in which the strain was applied in the $y$-direction since they are similar to the ones for the $x$-direction. With a rapid glimpse of Figure \ref{fig:mdstress}, we can note that all PG membranes present similar elastic behavior and fracture pattern. These membranes have just one stage of elastic deformation and the critical strain increases proportionally with the nanopore diameter from 22.33\% up to 33.74\%. After these strain thresholds, the PG membranes go abruptly from elastic to a completely fractured regime, which is characteristic of a brittle behavior. The larger is the nanopore diameter the greater is the tendency of a PG membrane in collapsing on parallel conjugated polymer chains during the stretching process. This trend contributes to diminish substantially the size of the membranes in the unstretched direction. The VM stress is uniformly distributed/accumulated on the membranes and for high values, it results in an abrupt rupture and fast crack propagation. The VM stress --- or equivalently the potential energy gain --- is stored in the carbon-carbon bonds almost parallel to the applied strain. Importantly, these bonds are the first to break, and two separated/unstressed pieces of PG lattices are obtained subsequently the membranes rupture. Other reactive MD simulations in the literature carried out using the ReaxFF \cite{desousa_MRSADV} and AIREBO \cite{felix_JPCC} potentials have shown that graphene sheets present a similar fracture process for a uniaxial stretching.

\begin{figure*}[pos=t]
\centering
\includegraphics[width=0.9\linewidth]{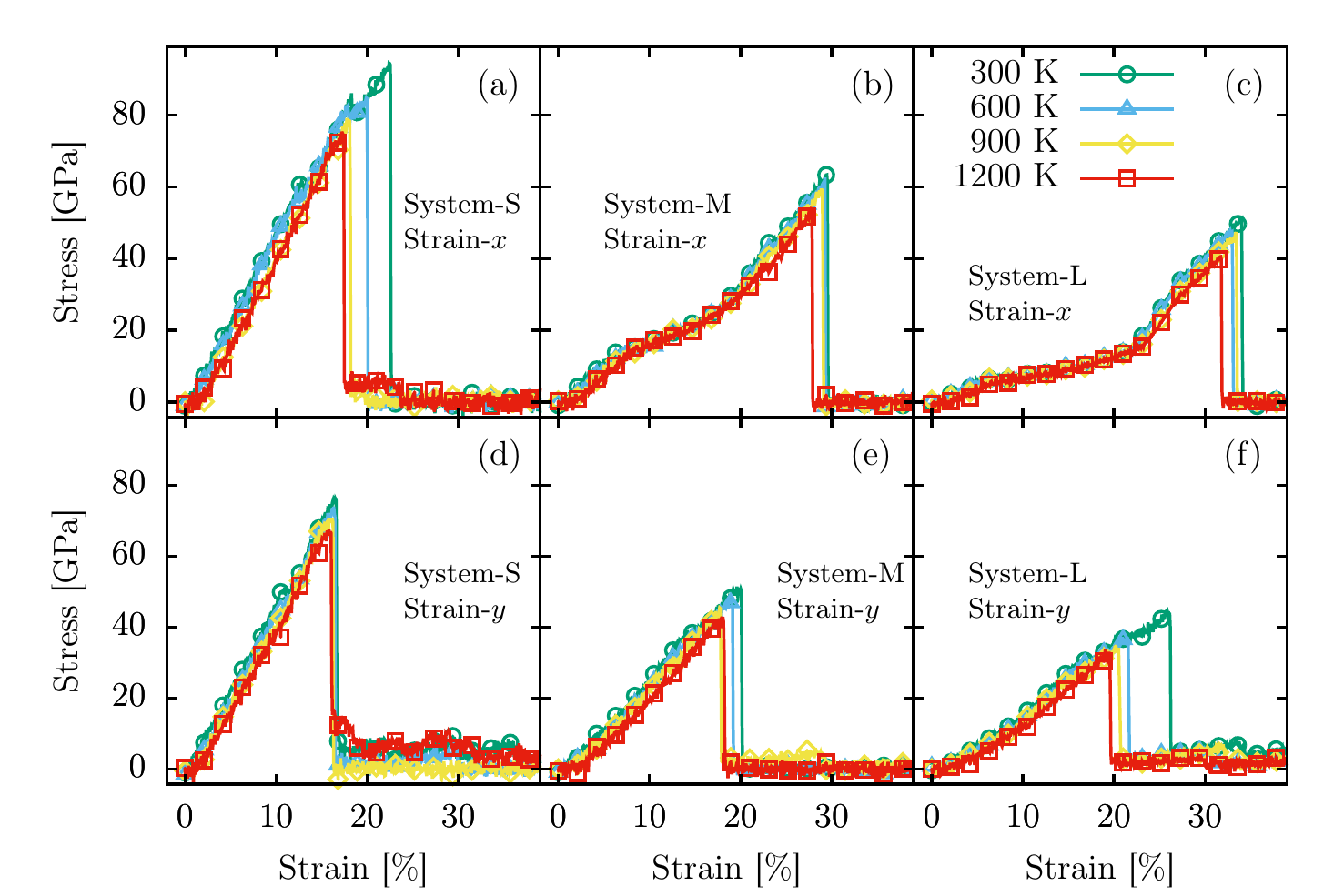} 
\caption{Calculated stress-strain curves for all the PG membranes when subjected to temperature regimes ranging from 300K to 1200K, considering a uniaxial strain applied in both $x$ and $y$ directions. The membranes were stretched at a constant rate until total rupture.}
\label{fig:sscurves}
\end{figure*}

Finally, we present in Figure \ref{fig:sscurves} the calculated stress-strain curves for all PG membranes when subjected to temperature regimes ranging from 300K to 1200K, considering a uniaxial strain applied in both $x$ and $y$ directions. As can it be perceived, the membranes were stretched at a constant rate until total rupture. The stress-strain curves shown the common regions that take place when nanostructures are stressed \cite{desousa_RSCADSV}: (I) a quasi-linear elastic regime is observed up to a maximum stress value (the ultimate strength ($US$)), (II) then the stress value drops abruptly to zero (or close to it) when the PG membranes start to fracture until they ultimately break (III) for a critical fracture strain ($FS$). Remarkably, the $US$ for PG complete fracture was noted to be piratically temperature independent. Instead, it depends only on the nanoporous diameter. The $US$ values are slightly higher for the strain in the $x$-direction. This trend occurs since in $x$-direction there are much more carbon-carbon parallel bonds to the applied strain, that share the accumulated stress than in $y$-direction. For the strain in the $y$-direction, the stress is stored in fewer carbon-carbon bonds, causing the membrane fracture for smaller $US$ values. In Figure \ref{fig:sscurves} one can note that the fracture strains range between 15\%--34\% by applying a uniaxial loading in both plane directions for temperatures up to 1200K. The lowest fracture strain was obtained for System--L at 1200K (15.82\%). As expected, increasing the temperature to 1200K, there is a decrease in the critical tensile strain values for all PG membranes (see \ref{table:elasprop}). Moreover, the highest Young's Modulus ($Y_M$) was obtained for System--S at 300K (471.01 GPa). As also expected, these results indicate that the $n$-Benzo-CMPs with small nanopore diameters are stiffer than the ones with large pores, which is denoted by the higher $Y_M$ as well as lower $FS$ values presented by System--L when contrasted to the other cases simulated here. Importantly, Table \ref{table:elasprop} summarizes the elastic properties ($Y_M$, $FS$, and $US$) that were obtained by fitting the stress-strain curves for the $n$-Benzo-CMPs cases investigated here.

\begin{table*}[]
	\begin{tabular}{clccclccc}
		\hline
		\multicolumn{1}{c}{\multirow{3}{*}{Temperature {[}K{]}}} & \multicolumn{1}{l}{} &                                                                                                        \multicolumn{7}{c}{System-S}                                                                                                        \\ \cline{3-9}
		                  \multicolumn{1}{c}{}                   & \multicolumn{1}{l}{} &                                     \multicolumn{3}{c}{ Strain-$x$}                                      & \multicolumn{1}{l}{} &                                      \multicolumn{3}{c}{Strain-$y$}                                      \\ \cline{3-5}\cline{7-9}
		                  \multicolumn{1}{c}{}                   & \multicolumn{1}{l}{} & \multicolumn{1}{c}{$Y_M$ {[}GPa{]}} & \multicolumn{1}{c}{FS {[}\%{]}} & \multicolumn{1}{c}{US {[}GPa{]}} & \multicolumn{1}{l}{} & \multicolumn{1}{c}{$Y_M$ {[}GPa{]}} & \multicolumn{1}{c}{FS {[}\%{]}} & \multicolumn{1}{c}{US {[}GPa{]}} \\ \cline{1-1}\cline{3-5}\cline{7-9}
		                          300                            &                      &               471.01                &              22.40              &              94.50               &                      &               460.07                &              16.66              &              76.28               \\
		                          600                            &                      &               451.69                &              19.88              &              84.86               &                      &               428.55                &              16.31              &              73.98               \\
		                          900                            &                      &               350.16                &              17.78              &              79.24               &                      &               396.75                &              16.03              &              70.71               \\
		                \multicolumn{1}{c}{1200}                 & \multicolumn{1}{l}{} &     \multicolumn{1}{c}{339.17}      &    \multicolumn{1}{c}{17.22}    &    \multicolumn{1}{c}{75.41}     & \multicolumn{1}{l}{} &     \multicolumn{1}{c}{369.35}      &    \multicolumn{1}{c}{15.82}    &    \multicolumn{1}{c}{67.43}     \\ 
	\end{tabular}
	
\begin{tabular}{clccclccc}
	\hline
	\multicolumn{1}{c}{\multirow{3}{*}{Temperature {[}K{]}}} & \multicolumn{1}{l}{}  &                                                                                                        \multicolumn{7}{c}{System-M}                                                                                                         \\ \cline{3-9}
	                  \multicolumn{1}{c}{}                   & \multicolumn{1}{l}{}  &                                     \multicolumn{3}{c}{ Strain-$x$}                                      & \multicolumn{1}{l}{}  &                                      \multicolumn{3}{c}{Strain-$y$}                                      \\ \cline{3-5}\cline{7-9}
	                  \multicolumn{1}{c}{}                   & \multicolumn{1}{l}{}  & \multicolumn{1}{c}{$Y_M$ {[}GPa{]}} & \multicolumn{1}{c}{FS {[}\%{]}} & \multicolumn{1}{c}{US {[}GPa{]}} & \multicolumn{1}{l}{}  & \multicolumn{1}{c}{$Y_M$ {[}GPa{]}} & \multicolumn{1}{c}{FS {[}\%{]}} & \multicolumn{1}{c}{US {[}GPa{]}} \\ \cline{1-1}\cline{3-5}\cline{7-9}
	                          300                            &                       &               236.04                &              29.47              &              63.40               &                       &               239.44                &              19.60              &              51.42               \\
	                          600                            &                       &               220.76                &              29.26              &              62.62               &                       &               208.17                &              18.83              &              48.12               \\
	                          900                            &                       &               183.66                &              28.84              &              59.56               &                       &               190.05                &              17.71              &              45.39               \\ 
	               \multicolumn{1}{c}{1200}                & \multicolumn{1}{l}{} &     \multicolumn{1}{c}{160.49}     &   \multicolumn{1}{c}{27.79}    &    \multicolumn{1}{c}{54.40}    & \multicolumn{1}{l}{} &     \multicolumn{1}{c}{131.33}     &   \multicolumn{1}{c}{18.01}    &    \multicolumn{1}{c}{42.82}    \\ 
\end{tabular}

\begin{tabular}{clccclccc}
	\hline
	\multicolumn{1}{c}{\multirow{3}{*}{Temperature {[}K{]}}} & \multicolumn{1}{l}{}  &                                                                                                        \multicolumn{7}{c}{System-L}                                                                                                         \\ \cline{3-9}
	                  \multicolumn{1}{c}{}                   & \multicolumn{1}{l}{}  &                                     \multicolumn{3}{c}{ Strain-$x$}                                      & \multicolumn{1}{l}{}  &                                      \multicolumn{3}{c}{Strain-$y$}                                      \\ \cline{3-5}\cline{7-9}
	                  \multicolumn{1}{c}{}                   & \multicolumn{1}{l}{}  & \multicolumn{1}{c}{$Y_M$ {[}GPa{]}} & \multicolumn{1}{c}{FS {[}\%{]}} & \multicolumn{1}{c}{US {[}GPa{]}} & \multicolumn{1}{l}{}  & \multicolumn{1}{c}{$Y_M$ {[}GPa{]}} & \multicolumn{1}{c}{FS {[}\%{]}} & \multicolumn{1}{c}{US {[}GPa{]}} \\ \cline{1-1}\cline{3-5}\cline{7-9}
	                          300                            &                       &               125.36                &              33.81              &              51.57               &                       &               137.31                &              25.83              &              25.74               \\
	                          600                            &                       &               115.06                &              32.76              &              47.82               &                       &               122.61                &              20.93              &              37.12               \\
	                          900                            &                       &                97.55                &              33.39              &              47.44               &                       &                95.28                &              20.16              &              34.56               \\ 
	               \multicolumn{1}{c}{1200}                & \multicolumn{1}{l}{} &     \multicolumn{1}{c}{81.65}      &   \multicolumn{1}{c}{31.71}    &    \multicolumn{1}{c}{40.59}    & \multicolumn{1}{l}{} &     \multicolumn{1}{c}{80.36}      &   \multicolumn{1}{c}{19.04}    &    \multicolumn{1}{c}{33.73}    \\ \hline
\end{tabular}
\caption{Elastic properties --- Young's Modulus ($Y_M$), Fracture Strain ($FS$), and Ultimate Strength ($US$) --- obtained by fitting the stress-strain curves (Figure \ref{fig:sscurves}) for the $n$-Benzo-CMPs cases investigated here.}
\label{table:elasprop}
\end{table*}

\section{Conclusions}
In summary, we have performed fully-atomistic reactive (ReaxFF) molecular dynamics simulations to investigate the elastic properties, fracture mechanism, and thermal stability of 2D porous $n$-Benzo-CMPs membranes with distinct nanopore architectures. For the sake of comparison, the results were contrasted with the ones for graphene sheets of similar dimensions.  We have considered three different H-passivated nanoporous diameters: small (3.45 \AA), medium (8.07 \AA), and large (11.93 \AA). At low temperatures (300K), the PG membranes present lattice configurations very close to their ground state arrangement. For temperatures about 3000K, PG membranes keep their integrity. However, thermal fluctuations lead these membranes to wrinkle due to large displacements presented by the atoms. Above 4500K, the PG membranes start to melt and their 2D $\pi$-conjugated polymer arrangement favors the formation of very short and linear atomic chains, which randomly combine carbon and hydrogen atoms. The initial configuration of the PG membranes vanishes completely for temperatures about 6000K, where only isolated atoms and small chain fragments that connect 2-10 atoms are observed resembling a gas-like phase of carbon atoms. For the microporous 2D $n$-Benzo-CMPs systems considered here, the melting points occur at 4660.07K, 4574.36K, and 4402.91 K, for the systems S, M, and L, respectively. Previous atomistic MD simulation has predicted a melting point for graphene about 5643K \cite{felix_JPCC}. 

The PG membranes under uniaxial tensile loading have presented similar elastic behavior and fracture pattern. They have just one stage of elastic deformation and the critical strain increases proportionally with the nanopore diameter from 22.33\% up to 33.74\%. After these critical strain values, the PG membranes go abruptly from elastic to a completely fractured regime. During the stretching process, the stress was uniformly accumulated on the membranes and for high values, it results in an abrupt rupture and fast crack propagation. The potential energy gain was stored in the carbon-carbon bonds almost parallel to the applied strain, and they were the first to break. Remarkably, the ultimate strength for PG complete fracture was noted to be temperature independent. Instead, it depends only on the nanoporous diameter. The lowest fracture strain was obtained for System--L at 1200K (15.82\%). As expected, increasing the temperature to 1200K, there is a decrease in the critical tensile strain values for all PG membranes. The highest Young's Modulus ($Y_M$) was obtained for System--S at 300K (471.01 GPa). 


\section*{Acknowledgements}
The authors gratefully acknowledge the financial support from Brazilian Research Councils CNPq, CAPES, and FAPDF and CENAPAD-SP for providing the computational facilities. L.A.R.J. gratefully acknowledges respectively, the financial support from FAP-DF  and CNPq grants 00193.0000248/2019-32 and 302236/2018-0. 

\printcredits
\bibliographystyle{unsrt}
\bibliography{cas-refs}

\end{document}